\documentclass[aip,amsmath,amssymb,reprint,]{revtex4-1}

\usepackage{graphicx}
\usepackage{dcolumn}
\usepackage{bm}
\usepackage{float}
\usepackage{color}
\usepackage[dvipsnames]{xcolor}
\usepackage{hyperref}
\usepackage{graphicx}
\usepackage{braket}
\usepackage{appendix}
\usepackage{chemmacros}
\usepackage{natbib}
\hypersetup{colorlinks=true, citecolor=blue, urlcolor=blue, linkcolor=blue}
\usepackage[utf8]{inputenc}
\usepackage[T1]{fontenc}
\usepackage{mathptmx}
\usepackage{etoolbox}

\makeatletter
\def\@email#1#2{%
 \endgroup
 \patchcmd{\titleblock@produce}
  {\frontmatter@RRAPformat}
  {\frontmatter@RRAPformat{\produce@RRAP{*#1\href{mailto:#2}{#2}}}\frontmatter@RRAPformat}
  {}{}
}
\makeatother
\begin{document}

\preprint{AIP/123-QED}

\title{Induction of large magnetic anisotropy energy and formation of multiple Dirac states in SrIrO$_3$ films: Role of correlation and spin-orbit coupling}
% Force line breaks with \\
\author{Amit Chauhan}
\author{B. R. K. Nanda}%
 \email{nandab@iitm.ac.in}
\affiliation{Condensed Matter Theory and Computational Lab, Department of Physics, IIT Madras, Chennai-36, India}
\affiliation{Center for Atomistic Modelling and Materials Design, IIT Madras, Chennai-36, India}
\affiliation{Functional Oxides Research Group, IIT Madras, Chennai-36, India}

\date{\today}

\begin{abstract}
The 5$d$ transition metal oxides, in particular iridates, host novel electronic and magnetic phases due to the interplay between onsite Coulomb repulsion ($U$) and spin-orbit coupling (SOC). The reduced dimensionality brings another degree of freedom to increase the functionality of these systems. By taking the example of ultrathin films of SrIrO$_3$, theoretically we demonstrate that confinement led localization can introduce large magnetic anisotropy energy (MAE) in the range 2-7 meV/Ir which is one to two order higher than that of the traditional MAE compounds formed out of transition metals and their multilayers. Furthermore, in the weak correlation limit, tailored terminations can yield multiple Dirac states across a large energy window of 2 eV around the Fermi energy which is a rare phenomena in correlated oxides and upon experimental realization it will give rise to unique transport properties with excitation and doping.
\end{abstract}

\maketitle

%\section{\label{sec:Introduction}Introduction}
In 5$d$ transition metal oxides, the delicate interplay between spin-orbit coupling and electron-electron correlations leads to a plethora of unconventional quantum states including Dirac and Weyl semimetals \cite{Yan2017,Vafek2014}, Mott and axion insulators \cite{Pesin2010,Wan2012} and quantum spin liquids \cite{Baskaran2007,Takagi2019,Kenney2019}. As a consequence, intense theoretical and experimental research has been pursued in the $5d$ transition metal oxides to realize novel non-trivial exotic phases. The SOC varies as $Z^4$ with $Z$ as the atomic number and substantially affects the electronic and magnetic properties of this family of oxides. The onsite correlation strength $U$ reduces due to the extended nature of the 5$d$ wave functions. Hence, in $5d$ oxides, $U$ and SOC become comparable with other competing interactions such as Hund's coupling, electron hopping strength, etc. These competing energy scales give rise to rich electronic and magnetic phases in the space spanned by $U$ and SOC.\par
In this context, the family of iridates, has triggered both theoretical and experimental community to realize novel ground states arising from the competition between SOC and $U$. For example, the pyrochlore family of iridates with chemical formula A$_2$Ir$_2$O$_7$ (A = Y, Eu, etc.) exhibit topological semimetal and Mott insulating phases \cite{Wan2011,Krempa2013,Topp2018}. The honeycomb iridates with chemical formulae A$_2$IrO$_3$ (A = Na, Li, etc.) host unconventional magnetic and spin liquid phases \cite{Rau2014,Chaloupka2013,Kenney2019,Reuther2014}. The Ruddlesden-Popper series of strontium iridates, Sr$_{n+1}$Ir$_{n}$O$_{3n+1}$ exhibits Mott insulating phase with in-plane canted antiferromagnetic ground state for quasi-monolayer system with n = 1 and out-of-plane collinear antiferromagnetic state \cite{Nauman2017,Kim2008,Wantabe2013,Yan2015,Fujiyama2012} for quasi-bilayer system with n = 2. On the contrary, the three-dimensional perovskite SrIrO$_3$ (n = $\infty$) exhibits nonmagnetic Dirac semimetallic state \cite{Fujioka2017,Carter2012}.\par
While the bulk SrIrO$_3$ (SIO) and other low-spin $d^5$ perovskite oxides are studied extensively, their members in the reduced dimensions which introduce another degree of freedom in the form of confinement invite a few open questions and scopes that remain largely unexplored. These are: (i) Whether the Dirac semimetal state in compounds like SrIrO$_3$ and CaIrO$_3$ survive or with breakdown of translation symmetry in one direction other non-trivial topological states evolve. (ii) How the layer termination governs the quantum phases due to the presence of dangling states or lack of them. (iii) With correlation and strong SOC driven spin anisotropy in these systems, whether the reduced dimensionality gives rise to large magnetic anisotropy energy (MAE), emerging out of second-order effect of SOC, and if so, it will be a very rare phenomena in oxides and can revolutionize the development of nanoscale storage devices.\par
With regard to thin films, few intriguing experimental observations have been made. For instance, for SIO, a metal-to-insulator transition (MIT) has been observed by either decreasing the film thickness below 4 nm or by lattice mismatch between the film and the substrate \cite{Biswas2014,Groenendijk2017}. A nonlinear Hall effect, anisotropic magnetoresistance and ferromagnetism has been experimentally reported at low temperature \cite{Chaurasiya2021}. In a recent experimental work, the spin Hall angle was observed to increase with the film thickness, manifesting enhanced spin-to-charge conversion efficiency \cite{Everhardt2019,Wang2019}. However, the layer termination of these films are not known.\par
Motivated by these experimental observations in SIO films and lack of quantum many-body analysis of the results, in this letter, by carrying out SOC tunable DFT+$U$ calculations on prototype ultrathin SIO films, we explore the novel quantum phases in the $U$-SOC domain which is generic to low-spin perovskites oxides. Specific to SIO, we demonstrate the cause of MIT and report the possible existence of multiple Dirac states. Irrespective of layer termination, we show that, these ultrathin films exhibits large MAE which is one to two order higher than the traditional MAE materials like transition metals and their multilayers \cite{Huang2021,Trygg1995,Wilhelm2000}.\par
\begin{figure}
\includegraphics[angle=-0.0,origin=c,height=3.6cm,width=8.5cm]{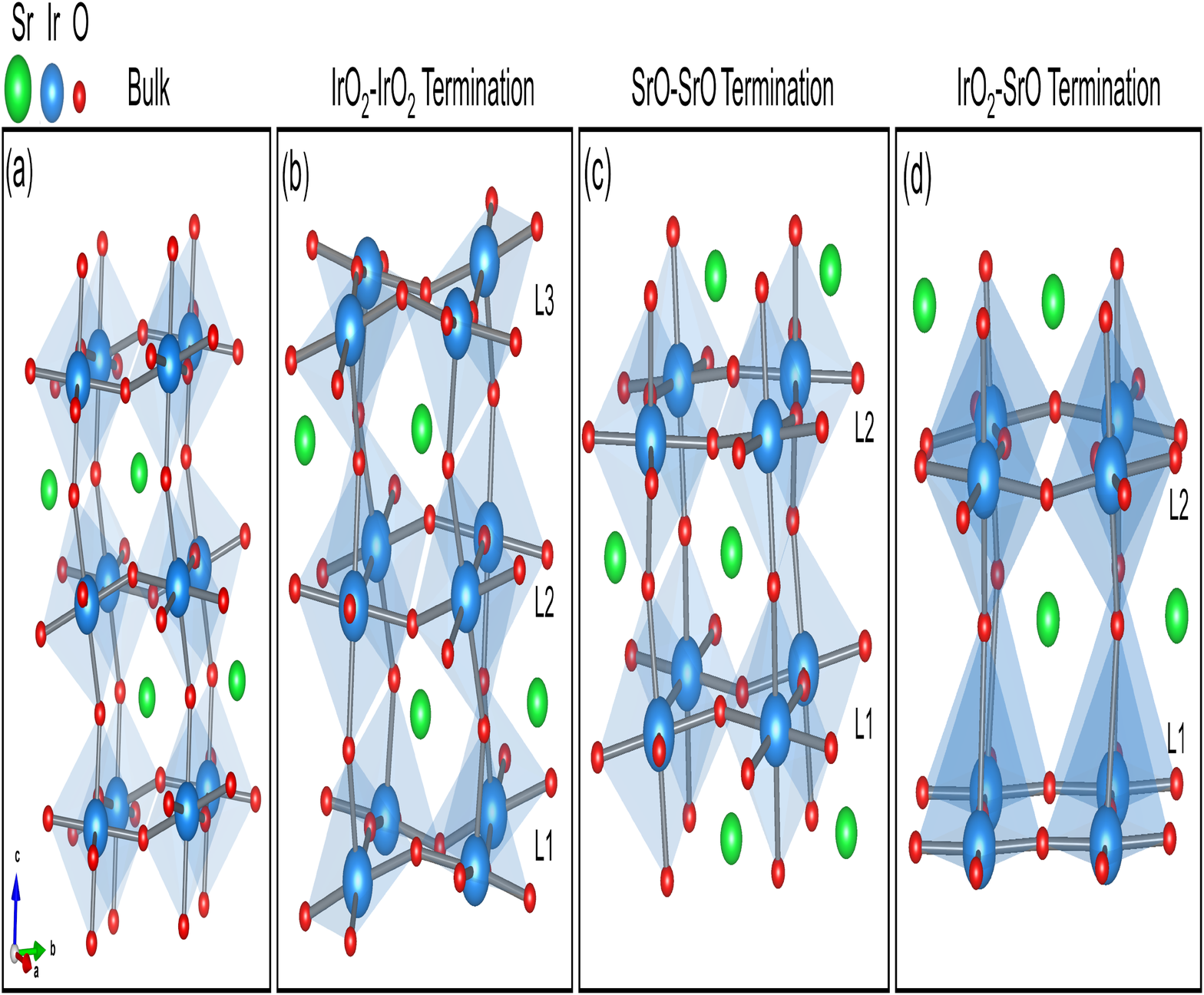}
\caption{(a) Crystal structure of bulk SrIrO$_3$. (b,c) Relaxed crystal structures for two symmetric films terminated by IrO$_2$ and SrO layers on both sides of the film, and (d) for one asymmetric film terminated by IrO$_2$ and SrO layers.}
\label{fig1}
\end{figure}

The crystal structures of bulk and ultrathin films of SIO are shown in Fig. \ref{fig1}. Due to distortion of octahedra, the bulk SIO crystallizes in an orthorhombic pervoskite crystal structure with Pbnm (62) space group which emerges out of a $\sqrt{2}a_\mathrm{0} \times \sqrt{2}a_\mathrm{0} \times 2a_\mathrm{0}$ supercell geometry with $a_\mathrm{0}$ as the ideal cubic lattice parameter. The experimental lattice parameters are $a$ = 5.56 {\AA}, $b$ = 5.59 {\AA} and $c$ = 7.88 {\AA} \cite{Zhao2008}. Here, we have constructed few atomic layers thick films and designed two symmetric terminations, IrO$_2$-IrO$_2$ and SrO-SrO, and one asymmetric termination, IrO$_2$-SrO, grown along the [001] direction as shown in the Fig. \ref{fig1}. The first film has three IrO$_2$ layers while the other two films have two IrO$_2$ layers coupled to each other through the sandwiched SrO layers. A 15 {\AA} thick vacuum is placed to design the films and each of them  are optimized with the aid of pseudopotential based QUANTUM ESPRESSO simulation tool \cite{Giannozzi2009}. The optimized films are shown in Fig. \ref{fig1}. The further structural information is provided in the supplementary material.
The electronic structure calculations are carried out in the plane-wave basis using the PAW potentials \cite{Kresse1999,Blochl1994} as implemented in VASP \cite{Kresse1996}. The PBE-GGA was chosen for the exchange-correlation functional. The onsite correlation effect is incorporated via an effective onsite correlation parameter $U_\mathrm{eff}$ = $U-J$ via the rotationally invariant Dudarev approach \cite{Dudarev1998}. The SOC $\lambda$ is varied in units of $\lambda_\mathrm{0}$, where $\lambda_\mathrm{0}$ is the real SOC strength. The Brillouin zone integration are carried out using $8 \times 8 \times 1$ Monkhorost-Pack k-mesh. The kinetic energy cutoff for plane-wave basis set was chosen to be $400$ eV.\par
\begin{figure}
\centering
\includegraphics[angle=-0.0,origin=c,height=6.5cm,width=7.5cm]{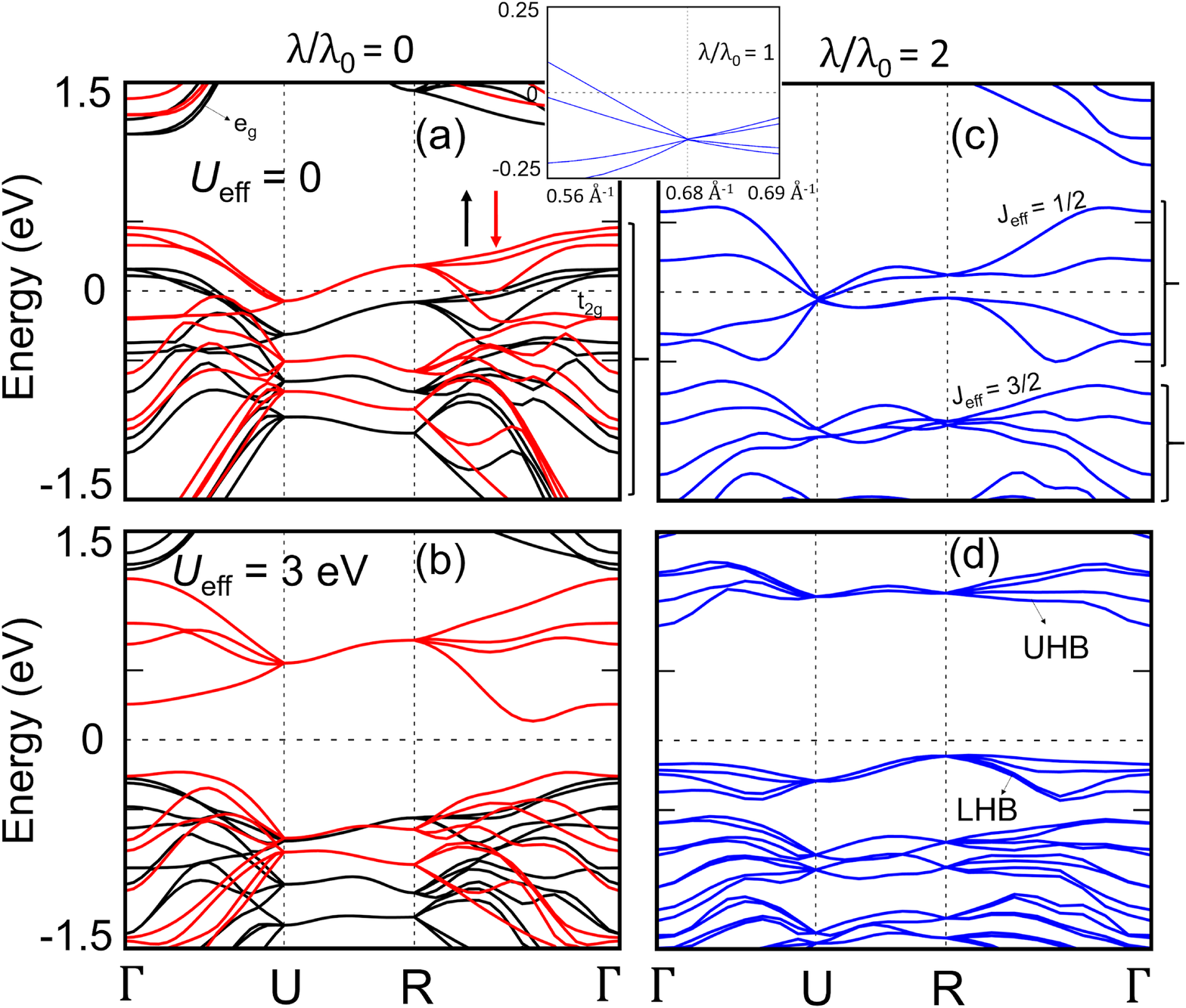}
\caption{Evolution of bulk electronic structure as a function of $U_{eff}$ and SOC. (a,b) Enhanced spin splitting leading to metal-insulator transition as the correlation strength increases. (c,d) formation of DSM phase with SOC and J$_{eff}$ = 1/2 Mott insulating phase with both SOC and $U$.}
\label{bulk_bands}
\end{figure} 
To provide a background, we will briefly present the electronic structure of bulk SIO through Fig. \ref{bulk_bands}. The details can be found in one of our earlier study \cite{Chauhan2021}. In its ground state, SIO exhibits Dirac semimetal (DSM) phase \cite{Chauhan2021,Carter2012}. However, the competition between correlation $U$ and SOC can alter the electronic structure significantly as demonstrated in the Fig. \ref{bulk_bands}. For $\lambda/\lambda_0$ = $U_{eff}$ = 0, a soft ferromagnetic metallic state, governed by crystal field split led t$_{2g}$-e$_{g}$ manifold, stabilizes. With onsite Coulomb repulsion, the enhanced spin splitting create a gap and stabilizes the system in ferromagnetic insulator state (see Fig. \ref{bulk_bands} (b)).
With finite SOC, real-spin states of the $t_\mathrm{2g}$ manifold mixes and leads to the formation of spin-orbit entangled $J_\mathrm{eff}$ = 1/2 and $J_\mathrm{eff}$ = 3/2 states. Due to the $d^5$ valence state of Ir, the $J_{eff}$ = 3/2 states are completely occupied whereas the $J_{eff}$ = 1/2 states are half-filled and merge at the high symmetric k-point U to create a Dirac node (see Fig. \ref{bulk_bands} (c)). With both $U$ and SOC, the $J_\mathrm{eff}$ = 1/2 doublets split into lower Hubbard and upper Hubbard subbands (LHB and UHB) to create a gap (see Fig. \ref{bulk_bands} (d)), leading to the formation of a canted antiferromagnetic insulating state.\par
\begin{figure*}
\centering
\includegraphics[angle=-0.0,origin=c,height=10cm,width=15cm]{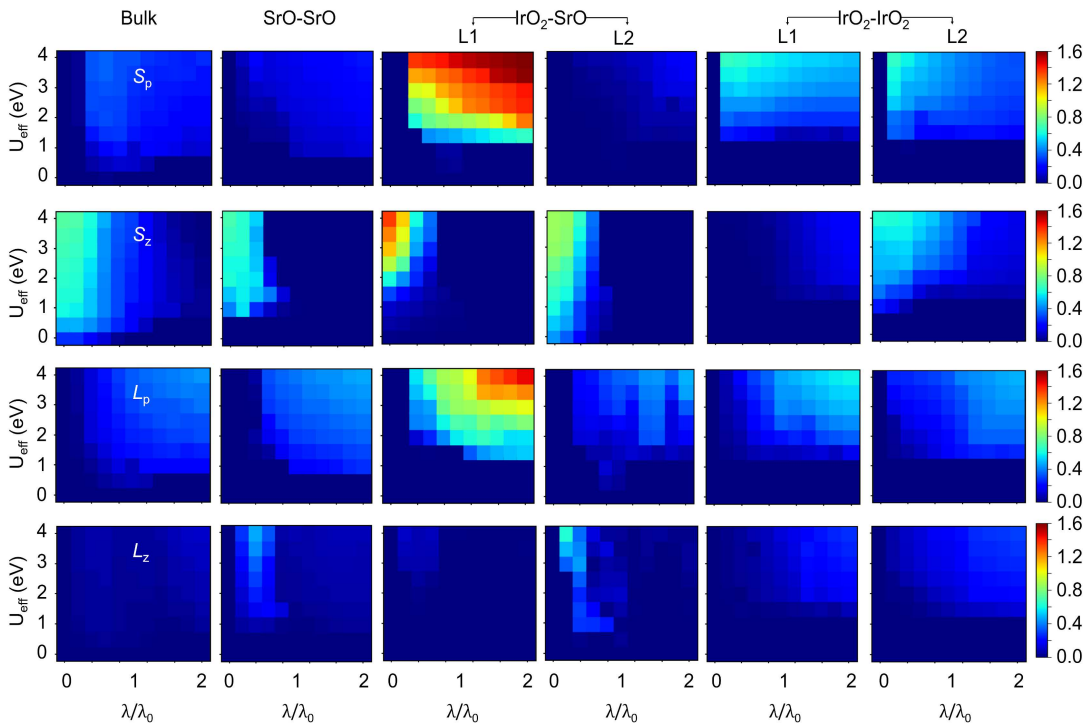}
\caption{Layer resolved spin and orbital intensity maps for bulk (first coloumn), SrO-SrO termination (second coloumn), IrO-SrO termination (third and fourth coloumn) and IrO$_2$-IrO$_2$ termination (fifth and sixth coloumn), respectively. The layer-1, layer-2 in SrO-SrO termination and layer-1, layer-3 in IrO$_2$-IrO$_2$ termination exhibits similar behaviour, hence, not explicitly shown here.}
\label{intensity}
\end{figure*}
The low-spin $d^5$ perovskites including iridates are capable of exhibiting eight different electronic and magnetic phases \cite{Chauhan2021}. However, if we restrict to the experimental orthorhombic structure, overall five phases are observed which include collinear ferromagnetic metal (FM), collinear ferromagnetic insulator (FI), Dirac semimetal (DSM), canted antiferromagnetic metal (CAFM) and canted antiferromagnetic insulator (CAFI) are obtained in the $U$-SOC domain. The layer resolved intensity maps for spin planar/normal (S$_p$/S$_z$) and orbital planar/normal (L$_p$/L$_z$) components for bulk are shown in the left most coloumn of Fig. \ref{intensity}. In the weak SOC regime ($\lambda$/$\lambda_o$ $\leq$ 0.2), only collinear FM and FI phases exist with negligible S$_p$ component.
For the intermediate range (0.4 $\leq$ $\lambda$/$\lambda_0$ $\leq$ 1.4) and low $U_{eff}$ values, the magnetic state weakens with vanishing S$_p$ and S$_z$ components and it leads to the formation of nonmagnetic DSM phase. With increasing $U_{eff}$ the effect of SOC gets enhanced as it favours the non-coplanar spin arrangement.  For the large $\lambda$/$\lambda_0$ regime ($\lambda$/$\lambda_0$ $\geq$ 1.4) there is a transition from the non-coplanar to pure co-planar spin arrangement in the $ab$-plane.\par
The pseudo mono and bilayer members of the iridate family, e.g. Sr$_2$IrO$_4$ and Sr$_3$Ir$_2$O$_7$, exhibit canted in-plane and collinear out-of-plane antiferromagnetic state through a spin-flop transition to indicate that reduced dimensionality has significant influence on the electronic and magnetic structure. With the present state-of-the art deposition techniques as well as innovative synthesis methods - such as by using lift-off and transfer methods or by etching of sacrificial water soluble layers \cite{Lu2016,Gu2020}- it is possible to artificially reduce the dimension and produce freestanding oxide films.
This inspires us to examine the formation of quantum phases in few atomic layer thick SIO freestanding films which are already discussed and shown in Fig. \ref{fig1}.\par
\begin{figure*}
\centering
\includegraphics[angle=-0.0,origin=c,height=11cm,width=14cm]{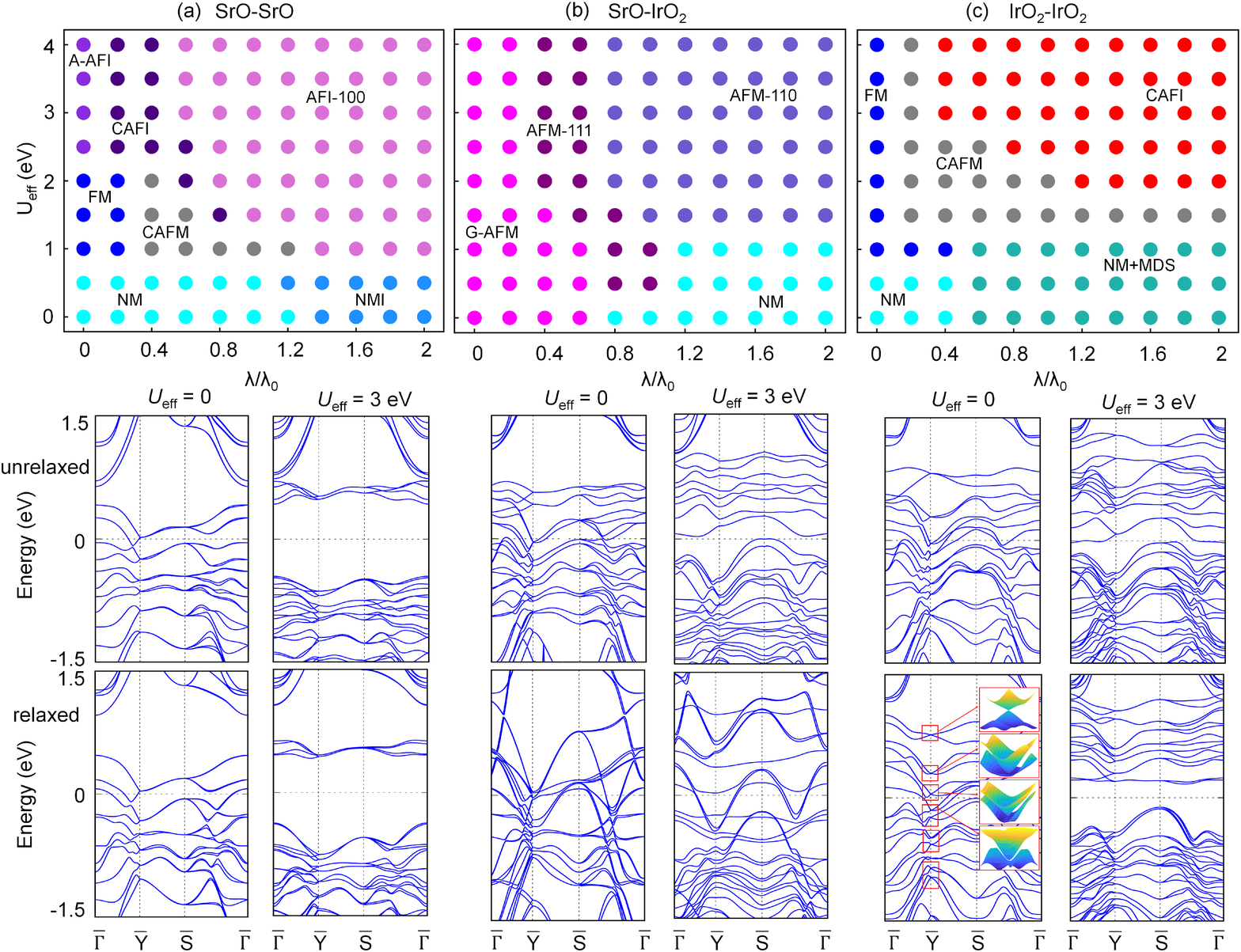}
\caption{Upper panel: Electronic and magnetic phase diagram of slabs terminated with (a) SrO layers (SrO-SrO), (b) SrO and IrO$_2$ layers (SrO-IrO$_2$), and (c) IrO$_2$ layers as a function of $U$ and SOC. Lower panel: The corresponding unrelaxed and relaxed band structures evaluated at $\lambda$/$\lambda_0$ =  1 with $U_{eff}$ = 0 and $U_{eff}$ = 3 eV, respectively.}
\label{phases}
\end{figure*}

% \textit{SrO-SrO and SrO-IrO$_2$ terminations}: 
The computed phase diagrams for SrO-SrO and SrO-IrO$_2$ terminations are shown in Fig. \ref{phases}. Let us first discuss SrO-SrO termination. We find stabilization of four new phases in addition to the five phases observed in the bulk \cite{Chauhan2021} and these are: nonmagnetic insulator (NMI), collinear A-type antiferromagnetic insulator (A-AFI), canted antiferromagnetic insulator (CAFI) with A-type magnetic ordering of spins along the c-axis and an in-plane antiferromagnetic insulating state with G-type spin ordering along the $\hat{x}$ direction (AFI-100).\par
In the weak SOC regime ($\lambda$/$\lambda_o$ $\leq$ 0.2), a NM state stabilizes for weak $U_{eff}$ and transforms to the FM state for intermediate strength of the Coulomb repulsion with spins oriented along $\hat{z}$ as can be seen form spin intensity map of Fig. \ref{intensity}. Further, with increasing correlation strength, a transition from FM to A-AFI state occurs for higher $U_{eff}$. It may be noted here that in the bulk, no such transition occurs with increasing $U_{eff}$ and FM ordering of spins persists even for large values of $U_{eff}$.\par
In the intermediate SOC regime (0.4 $\leq$ $\lambda$/$\lambda_0$ $\leq$ 1.4), there is a very narrow domain in which CAFM and CAFI phases exists for intermediate and high $U_{eff}$ values (see Fig. \ref{phases}). The DSM phase vanishes in the weak $U_{eff}$ regime and a trivial NM/NMI state stabilizes. It has an interesting consequence in the spin decomposition. If we compare the variation in $S_z$ between bulk and film (see Fig. \ref{intensity}), it vanishes rapidly for the latter with SOC and stabilizes the AFI-100 phase, manifesting the enhanced effect of SOC with the reduced dimensionality. Infact, the MAE ($\approx$ $E^x$-$E^z$), which will be discussed later, is estimated to be $\approx$ -2.7 meV/Ir at $\lambda$/$\lambda_0$ = 1 and $U_{eff}$ = 3 eV indicating that the in-plane magnetization is strongly favoured by SOC. The canted phases appear in a very narrow domain. In the large SOC regime, either PMI phase exists for low $U_{eff}$ values or AFI-100 phase for intermediate and high $U_{eff}$ values. The planar spin and orbital components (see Fig. \ref{intensity}) dominates in the strong SOC regime.\par
The aforementioned observation can be correlated to the pseudo bilayer compound Sr$_3$Ir$_2$O$_7$ and the heterostructure 2SIO/1STO for validation. The former exhibits collinear out-of-plane antiferromagnetic state with negligible tilt angle ($\theta_{t}$) $\approx$ 179.5$^\circ$ whereas the latter exhibits c-axis canted antiferromagnetic state with finite $\theta_{t}$ $\approx$ 172$^\circ$ \cite{Meyers2019}. Interestingly, earlier theoretical investigations suggests that with increasing tilting the spin ordering in the bilayer heterostructure turns coplanar ($ab$ plane) \cite{Mohopatra2019}. Concurring to the above observations, the present study on SrO-SrO termination, where the $\theta_{t}$ $\approx$ 164$^\circ$ (see Fig. 1 in SI), stabilizes the coplanar spin ordering with $U_{eff}$ close to and above 1 eV.\par
The phase diagram of four layer thick asymmetric slab (SrO-IrO$_2$) is shown in Fig. \ref{phases}. Due to a large scale structural distortion, we observe metallic phases in the entire $U$-SOC domain. In the weak SOC regime, a collinear G-type antiferromagnetic metal (G-AFM) state stabilizes independent of the $U_{eff}$ value. In the intermediate SOC regime, the collinear AFM-111 state stabilizes, whereas the coplanar AFM-110 state stabilizes for higher SOC strength. The layer resolved spin intensity maps for IrO$_2$-SrO termination of Fig. \ref{intensity} clearly show that the S$_p$ and L$_p$ components of layer-1 (terminating layer) dominates as compared to layer-2 with quite high S$_p$ component value. This is due to the breakdown of the low-spin picture of the $d^5$ electronic configuration in the octahedral symmetry as the large scale distortion leads to the formation of nearly square planar complex in layer-1.\par
% \textit{IrO$_2$-IrO$_2$ termination}:
The phase diagram for the slab terminated with IrO$_2$ layers is shown in Fig. \ref{phases}. In the weak SOC regime, a NM state transforms to FM state. However, unlike the bulk case, we did not observe a metal-to-insulator transition with increasing $U_{eff}$. In the intermediate SOC regime, while correlation remains weak, a nonmagnetic phase with multiple Dirac states (NM+MDS) protected by nonsymmorphic symmetry form at the high symmetric point $\bar{Y}$ over a large energy window in the vicinity of Fermi energy ($E_{F}$) as can be seen from the band structures of Fig. \ref{phases}(c). The system retains the Dirac states even for finite but weak correlation ($U_{eff}$ $\approx$  1 eV) as can be seen from Fig. 2 of supplementary information. Experimental realization of such a phase will give rise to interesting electron and spin transport properties upon carrier doping. It is interesting to note that unlike the SrO-SrO and SrO-IrO$_2$ terminated films, the noncoplanar spin ordering persists here even for large values of $U_{eff}$ and $\lambda$/$\lambda_0$.\par
The experimental studies on unstrained SIO films show metallicity up to 4 nm and below it the system becomes insulating \cite{Biswas2014}. On compressed films, the metal-to-insulator transition has been observed between 5 to 10 nm \cite{Everhardt2019}. While the layer terminations of these films are not known, the phase diagrams developed in this work provides the theoretical insight to the probable cause of formation of these electronic phases.\par
% \textit{Magnetic anisotropy energy}:
The presence of strong SOC ($\approx$ 0.43 eV) and tunable spin anisotropy make the iridate films ideal to induce large MAE which is crucial in the application domains of spintronics. The magnetic easy axis as well as the MAE magnitude can be found by calculating the difference in the second-order induced SOC energies for in-plane ($E^x$) and out-of-plane ($E^z$) spin orientations (MAE $\approx$ $E_{SOC}^x$ - $E_{SOC}^z$).\par 
For the films with IrO$_2$-IrO$_2$, SrO-SrO, and IrO$_2$-SrO terminations, we computed the MAE to be -2, -2.7 and -7 meV/Ir, for $\lambda$/$\lambda_0$ = 1 and $U_{eff}$ = 3 eV, respectively. The present study, though use SrIrO$_3$ as a prototype, is aimed towards the class of low-spin $d^5$ perovskites. The phase diagrams show that the films become strongly magnetic for $U_{eff}$  $>$ 2 eV. Therefore, to qualitatively find the MAE, we have used $U_{eff}$ = 3 eV. The order of MAE is expected to remain unchanged with higher $U_{eff}$. The MAE magnitudes are one to two order higher than the traditional MAE materials like transition metals and their multilayers. For example, among the ferromagnetic $3d$ metals Fe, Co and Ni, Co exhibits highest MAE of $\approx$ 0.06 meV/atom \cite{Trygg1995}. On the other side, the Fe/V, Ni/Pt and Co/Au multilayers exhibits MAE of $\approx$ 0.002, 0.02 and 0.16 meV/atom \cite{Wilhelm2000}. It is very rare that oxide films possess large MAE. A very recent theoretical study on SrRuO$_3$ thin films report a MAE of $\approx$ 0.62 meV/Ru \cite{Huang2021}. Also, for comparison, the MAE for other known oxides is summarized in Table-I.\par
\begin{figure}[H]
\centering
\includegraphics[angle=-0.0,origin=c,height=5cm,width=6cm]{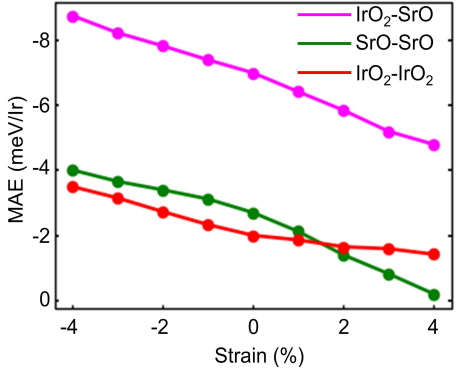}
\caption{Variation in magnetic anisotropy energy with strain for IrO$_2$-SrO, SrO-SrO and IrO$_2$-IrO$_2$ terminations.}
\label{Fig5}
\end{figure}

Epitaxial strain is often used as an external agent to tune the film properties. To observe the effect of it on MAE, we have applied tensile and compressive strain on the films and the results are shown in Fig. \ref{Fig5}. The MAE sign remains negative for both tensile as well as compressive strain, suggesting that the easy axis lies in the plane and is robust against external perturbation. Independent of the layer termination, the MAE magnitude increases linearly with compression and decreases with expansion.\par
\begin{figure}[H]
\includegraphics[angle=-0.0,origin=c,height=3.5cm,width=8.5cm]{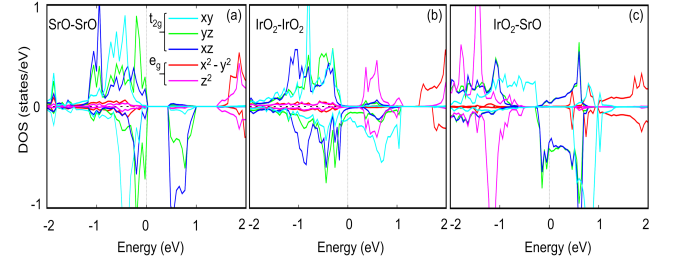}
\caption{Orbital and spin resolved density of states calculated within GGA+$U$ with $U_{eff}$ = 3 eV for (a) SrO-SrO, (b) IrO$_2$-IrO$_2$ and, (c) IrO$_2$-SrO terminations.}
\label{DOS}
\end{figure}

\begin{table}
\caption{Magnetic anisotropy energy of transition metal (TM) oxides in units of meV per TM atom.}
\begin{tabular}[t]{|p{2cm}|c|}
\hline
TM oxides&MAE(meV/TM atom)\\ \hline
SrIrO$_3$(present work)&2-7 \\ \hline
SrRuO$_3$&0.62\\ \hline
Fe$_2$O$_3$&0.001 \\ \hline   
NiO&0.015\\ \hline
MnO&0 \\ \hline
FeO&1\\ \hline
CoO&1 \\ \hline 
\end{tabular}
\end{table}
A qualitative analysis of MAE can be obtained from the SOC contribution to the second-order perturbation correction. In terms of angular momentum, it is expressed as \cite{Freeman1993}
\begin{align*}
    MAE \simeq\lambda^2 \sum_{u^\sigma,o^{\sigma'}} \frac{\mid\bra{u^\sigma}\hat{L_z}\ket{o^{\sigma'}}\mid^2-\mid\bra{u^\sigma}\hat{L_x}\ket{o^{\sigma'}}\mid^2}{\epsilon_{u^\sigma}-\epsilon_{o^{\sigma'}}}\alpha,
\end{align*}
where $\lambda$ is the SOC strength and $\alpha$ = $2\delta_{\sigma,\sigma'}-1$. u$^\sigma$ (o$^{\sigma'}$) denotes the unoccupied (occupied) states in spin ($\sigma$, $\sigma^\prime$  = {$\uparrow$, $\downarrow$}) channels. $\epsilon_{u^\sigma}$ - $\epsilon_{o^{\sigma'}}$ is the energy difference between the band centers of unoccupied and occupied states. The non-vanishing matrix elements of the above equation that contribute to MAE are: $\bra{xz}\hat{L_z}\ket{yz} = 1$, $\bra{x^2 - y^2}\hat{L_z}\ket{xy} = 2$, $\bra{z^2}\hat{L_x}\ket{yz} = \sqrt{3}$, $\bra{xy}\hat{L_x}\ket{xz} = 1$ and $\bra{x^2-y^2}\hat{L_x}\ket{yz} = 1$. The spin and orbital resolved density of states (see Fig. \ref{DOS}) qualitatively  estimates the band centers and hence the energy difference $\epsilon_{u^\sigma}$ - $\epsilon_{o^{\sigma'}}$. For the SrO-SrO termination, the positive contribution to MAE comes from the matrix elements $\bra{yz,xz\downarrow}\hat{L_x}\ket{xy\uparrow}$ and  $\bra{xz\downarrow}\hat{L_z}\ket{yz\downarrow}$, whereas the negative contribution comes from the matrix elements $\bra{xz\downarrow}\hat{L_z}\ket{yz\uparrow}$ and $\bra{yz,xz\downarrow}\hat{L_x}\ket{xy\downarrow}$. Since all the above matrix elements have same magnitude, i.e. 1, whether the total MAE is negative or positive will depend on the denominator $\epsilon_{u^\sigma}$ - $\epsilon_{o^{\sigma'}}$.
As the denominator is found to be small for the negative matrix elements as compared to the positive ones, the SrO terminated film yields negative MAE. The terms involving the occupied t$_{2g}$ and unoccupied e$_g$ subbands are small due to large denominator. Also, the opposite spin terms among them nearly cancel each other to make negligible contribution to MAE. For the IrO$_2$-IrO$_2$ termination, the dominant contribution (negative) comes from the coupling between states in the down-spin channel, i.e., $\bra{xy\downarrow}\hat{L_x}\ket{yz,xz\downarrow}$. For the asymmetric termination, the dominant contribution (negative) originates from the coupling between localized z$^2$ states with yz and xz states ($\bra{yz,xz\downarrow}\hat{L_x}\ket{z^2\downarrow}$ and $\bra{yz,xz\uparrow}\hat{L_x}\ket{z^2\uparrow}$). The negative MAE for each of these three terminations implies that the magnetic easy axis lies in the plane and is independent of the layer termination.\par  
In summary, we have carried out first-principles based SOC tunable DFT+$U$ calculations to examine the effect of reduced dimensionality and layer termination in SrIrO$_3$ films by constructing few atomic layer thick slabs with symmetric and asymmetric layer terminations. Our analysis reveals that strong correlation effect can induce large magnetic anisotropy energy, 2-7 meV/Ir, in these films. For the slab terminated with IrO$_2$ layers, if the system has weak correlation, as in the case of bulk, a nonmagnetic phase possessing multiple Dirac states in a wide energy spectrum develops. Furthermore, the phase diagrams that we established in the $U$-SOC domain infers the possible formation of six unique magnetic configurations. It also provides the theoretical insight to when and why an insulating phase forms in these films as being reported in some of the earlier experimental studies.\par
The present study uncovers the key interaction mechanisms to establish different electronic and magnetic states and transition among them in a broad class of low-spin d$^5$ perovskites. Experimental validation will open-up new avenues to explore novel physics and applications in quantum condensed matter. The probable presence of large magnetic anisotropy energy make the thin films of these perovskites promising candidates for applications in magnetic memory and storage devices.\\

See the supplementary material for the information regarding the structural details of the bulk and ultra-thin films and electronic band structures for the IrO$_2$-IrO$_2$ terminated film with varying spin-orbit coupling.\\

The authors would like to thank HPCE, IIT Madras for providing the computational facility. This work is funded by the Department of Science and Technology, India, through grant No. CRG/2020/004330.\par

\section*{Data Availability}
The data that support the findings of this study will be available from the corresponding authors upon reasonable request.

\end{document}